\documentclass{prop2015v2}
\usepackage[english]{babel}
\usepackage{bbm,nicefrac}

\def\Z{\mathbb{Z}}
\def\R{\mathbb{R}}

\def\OR{\Omega\mathcal{R}}

\keywords{String theory, orbifold, D-brane, singularity, deformation, moduli stabilisation.}
\title{Deforming D-brane models on $\boldsymbol{T^6/(\Z_2 \times \Z_{2M})}$ orbifolds}
\author[I. Koltermann]{I. Koltermann\inst{1,}\footnote{Corresponding author;\; E-mail:~\textsf{kolterma@uni-mainz.de}}}
\author[M. Blaszczyk]{M. Blaszczyk\inst{1}}
\author[G. Honecker]{G. Honecker\inst{1}}
\address[1]{PRISMA Cluster of Excellence \& Institut f\"ur Physik  (WA THEP), Johannes-Gutenberg-Universit\"at, D-55099 Mainz}
\shortauthors{I. Koltermann et al.}

\begin{abstract}
We review the stabilisation of complex structure moduli in Type IIA orientifolds, especially on $T^6 / (\Z_2 \times \Z_6^\prime \times \OR)$ with discrete torsion, via deformations of $\Z_2 \times \Z_2$ orbifold singularities. While D6-branes in SO($2N$) and USp($2N$) models always preserve supersymmetry and thus give rise to flat directions, in an exemplary Pati-Salam model with only U($N$) gauge groups ten out of the 15 deformation moduli can be stabilised at the orbifold point.
\end{abstract}

\shortabstract

\begin{document}
\maketitle

\section{Introduction}

The usual backgrounds in string model building are generic Calabi-Yau threefolds or orbifolds as singular limits thereof. Both contain in general flat directions for the dilaton and geometric moduli. Our computations demonstrate that a large number of these flat directions can be stabilised~\cite{Blaszczyk:2015oia} without introducing closed string background fluxes~\cite{Grana:2005jc}, which implies that no severe back-reaction on the geometry occurs. In addition, our method is appropriate for calculating the tree-level value of gauge couplings by using periods over \textit{(special) Lagrangian}
(\textit{(s)Lag}) three-cycles, where previously identical couplings $g^{-2}_a \propto \int_{\Pi_a} |\Omega_3|$
 can be tuned via deformations to obtain phenomenologically acceptable values~\cite{Blaszczyk:2015oia,Blaszczyk:2014xla}.

\section{Deformations of Orbifold Singularities and Hypersurface Formalism}

For concreteness, we work with a factorisable six-torus for the $T^6/(\Z_2 \times \Z_{2M})$ orbifolds (with discrete torsion $\eta = -1$), on which we will perform \textit{deformations} of the $\Z_2 \times \Z_2$ singularities (contrary to \textit{blow-ups} for orbifolds with $\eta = +1$). The resulting exceptional three-cycles can be used for D6-brane model building. We will describe the orbifolds as $\bigl( T^6 / (\Z_2 \times \Z_2) \bigr) / \Z_M$, where ideally $\Z_M$ does not lead to additional exceptional three-cycles, but restricts the way how the $\Z_2 \times \Z_2$  singularities can be deformed. In particular, we will discuss the phenomenologically appealing $T^6 / (\Z_2 \times \Z_6^\prime \times \OR)$ orientifold on the SU$(3)^3$ lattice~\cite{Blaszczyk:2015oia}.

The basic three-cycles in this setup are {\it bulk} cycles, specified by a a product of toroidal one-cycles with integer-valued wrapping numbers, and {\it exceptional} cycles, where a two-cycle on a resolved $\Z_2^{(k)}$ singularity along $T^4_{(k)} \equiv T^2_{(i)} \times T^2_{(j)}$ is tensored with a one-cycle on $T^2_{(k)}$.
{\it  Fractional} cycles then consist of ($\nicefrac{1}{4}$ times) a bulk cycle passing through orbifold singularities and an appropriate contribution from the corresponding exceptional cycles. Only D6-branes wrapping so-called {\it sLag} 
three-cycles $\Pi$ with 
$\mathcal{J}_{1,1} \big|_{\Pi} = 0$, \,
$\Im\left(\Omega_{3}\right) \big|_{\Pi} = 0$, \,
$\Re\left(\Omega_{3}\right) \big|_{\Pi} > 0$
preserve $\mathcal{N}=1$ supersymmetry in four dimensions. While all aforementioned three-cycles are automatically \textit{Lag} 
(i.e.\ the first equation involving the K\"ahler form holds true), the two constraints containing the holomorphic volume form $\Omega_3$ have to be tested explicitly.

To deform the $T^6/(\Z_2 \times \Z_2)$ orbifold, a reformulation as hypersurface in an ambient toric space is useful. As a first step, one describes the two-torus as a hypersurface in the complex weighted projective space $\mathbbm{P}^2_{112}$ with homogeneous coordinates $(x, v, y)$. The elliptic curve is then the zero locus of a polynomial of degree four,
\begin{align*}
	f := -y^2 + F \stackrel{!}{=} 0 \,, \;\;\;\,
	F(x,v) = 4 \, v \, ( x-\epsilon_2 v) ( x- \epsilon_3 v ) (x - \epsilon_4 v) \,,
\end{align*}
where, due to the $\Z_2$ reflection $y \mapsto -y$, the fixed points are exactly the roots of $F(x,v)$. The anti-holomorphic orientifold involution is given by $\sigma_{\cal R}: (x,v,y) \mapsto (\overline x, \overline v,\overline y)$. 

The \textit{Lag} lines in untilted lattices ($\epsilon_4 < \epsilon_3 < \epsilon_2$, $\epsilon_i \in \R$) are then basically 
\includegraphics[height=12pt]{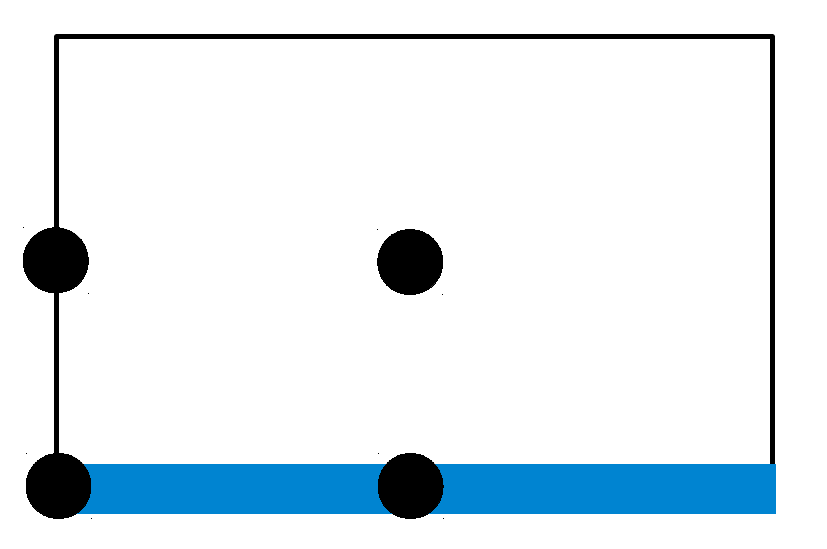} ,
\includegraphics[height=12pt]{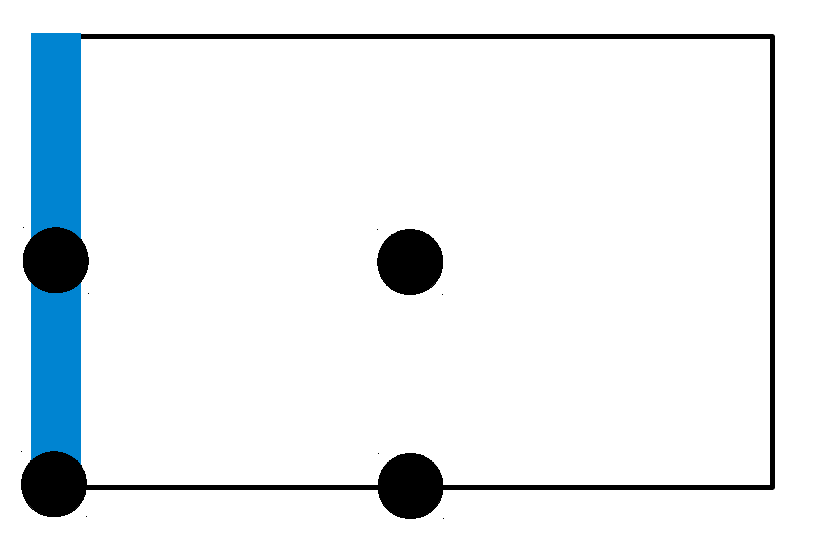} ,
\includegraphics[height=12pt]{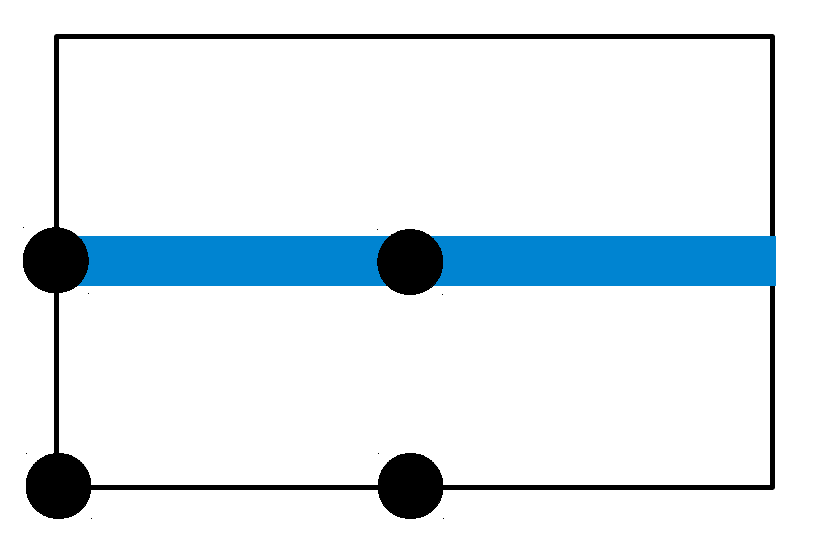} ,
\includegraphics[height=12pt]{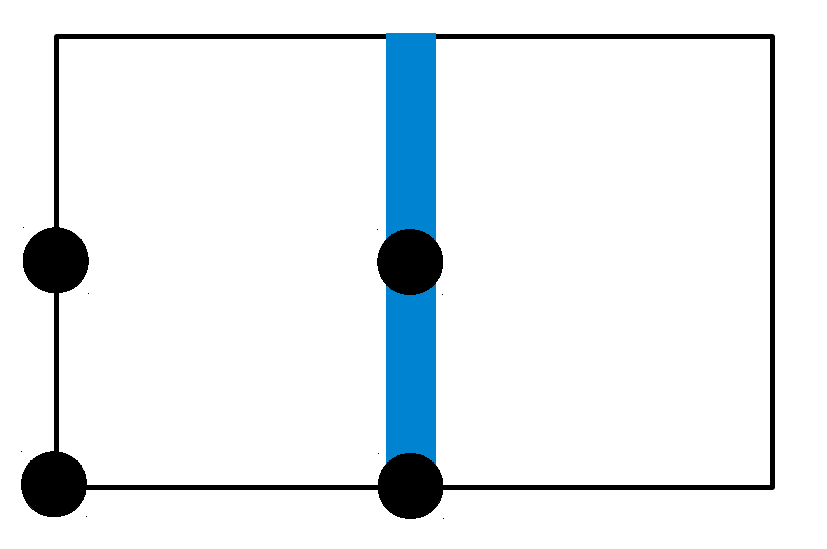},
and for tilted lattices ($\epsilon_2 = \overline\epsilon_4$, $\epsilon_3 \in \R$), which in particular include the hexagonal tori with their additional $\Z_3$ symmetry, we observe 
\includegraphics[height=12pt]{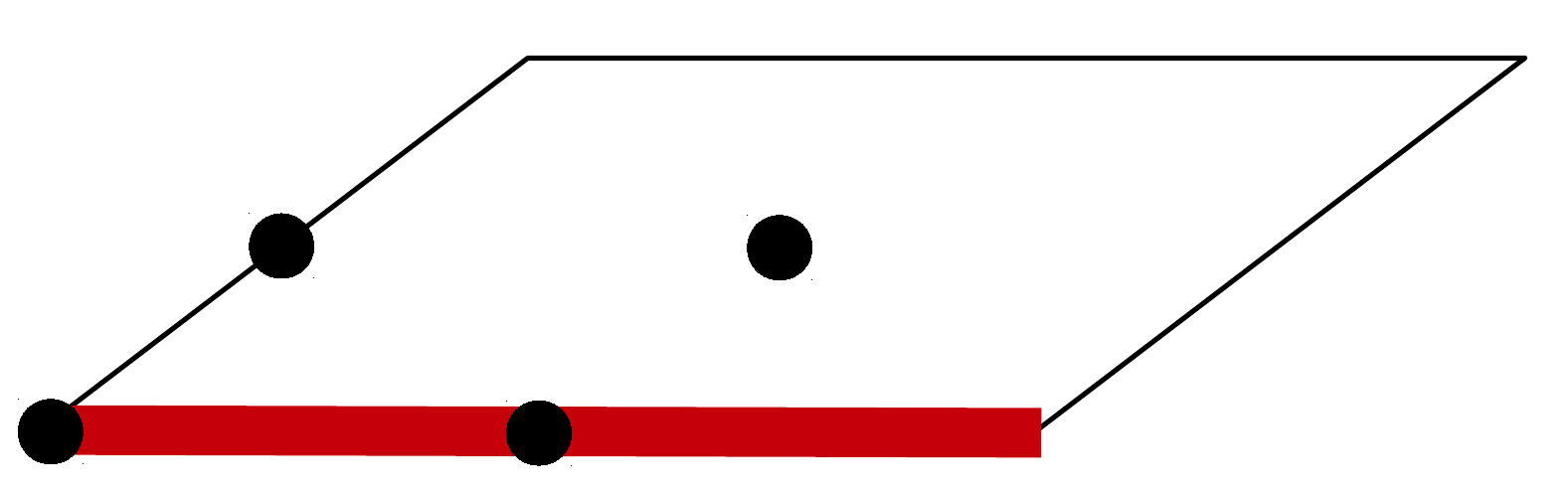} ,
\includegraphics[height=12pt]{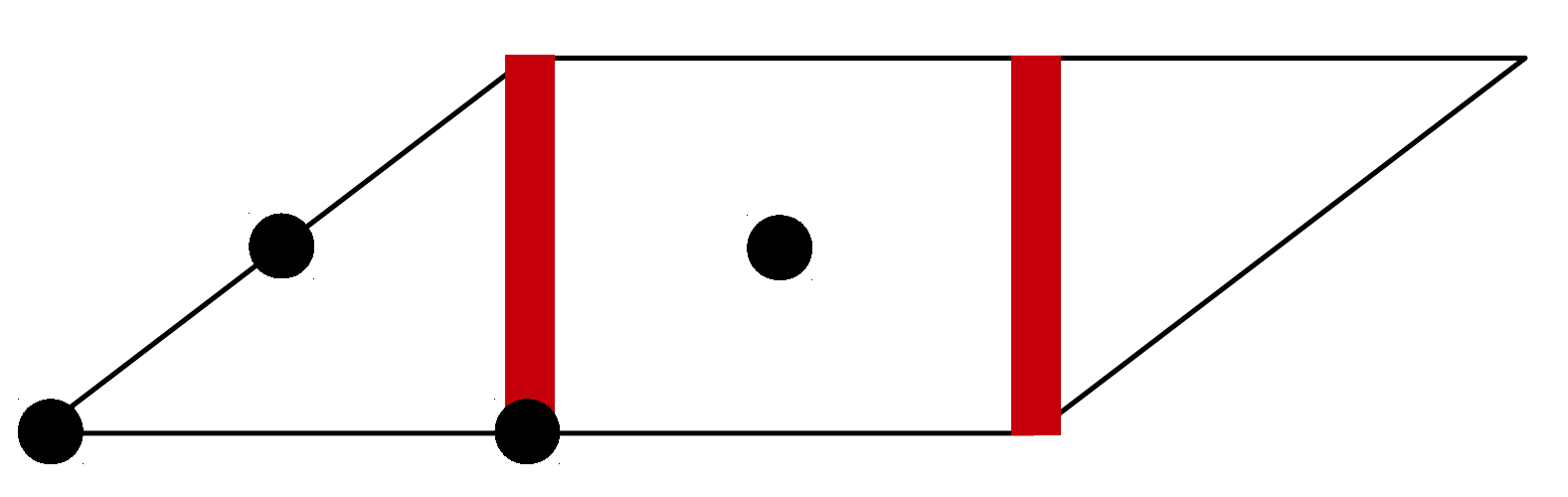} ,
\includegraphics[height=12pt]{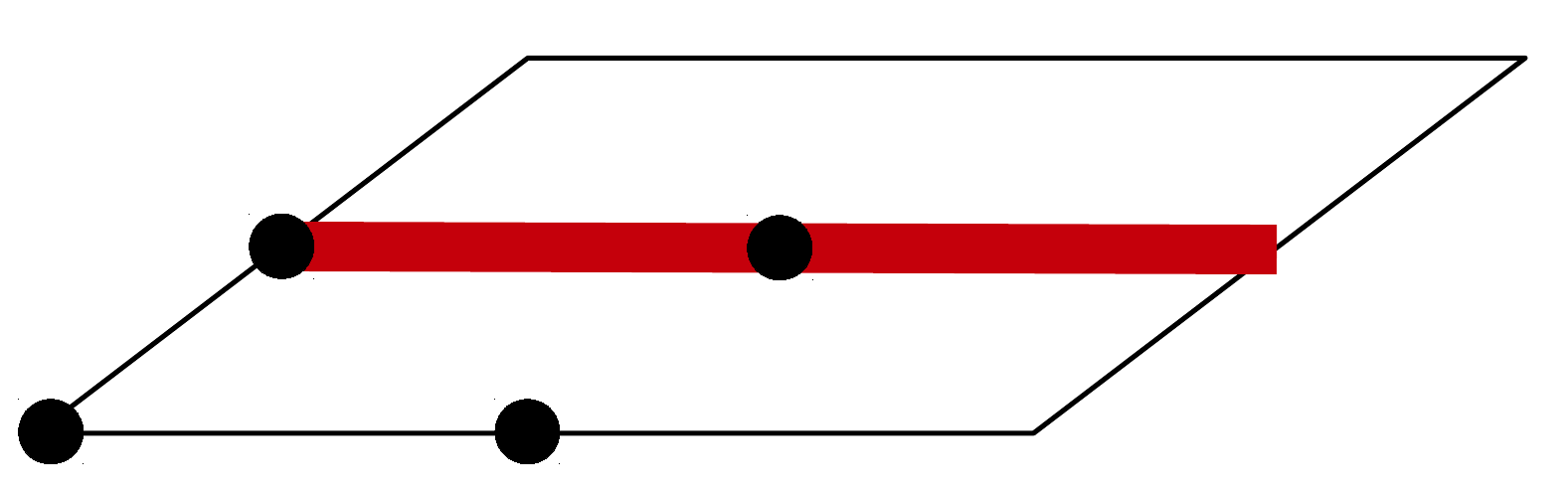} ,
\includegraphics[height=12pt]{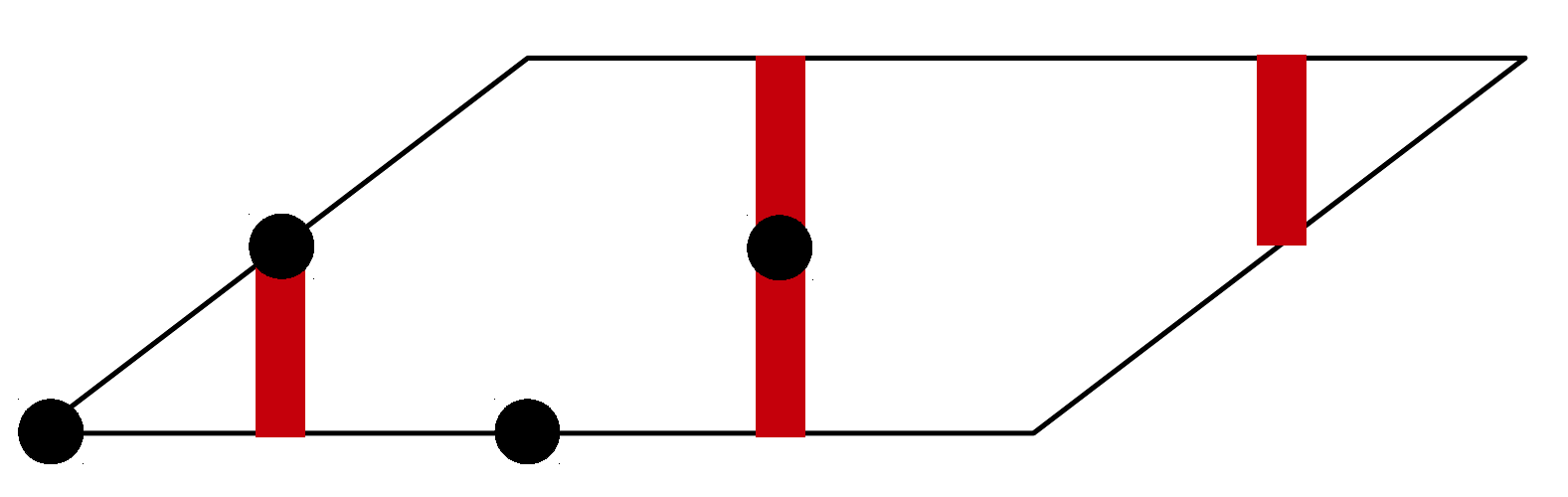}.

The $T^6/(\Z_2 \times \Z_2)$ orbifold and its deformation is described by the zero locus of
\begin{align*}
	f = -y^2 + F_1 F_2 F_3 - \sum_{i\neq j \neq k \neq i} \sum_{\alpha,\beta=1}^4
		\varepsilon^{(i)}_{\alpha\beta} \, F_i \, \delta F_j^{(\alpha)} \, \delta F_k^{(\beta)}
		 \,,
\end{align*}
where $F_i(x_i,v_i)$ encodes the complex structure of the two-torus $T^2_{(i)}$ and $\delta F_i^{(\alpha)}(x_i,v_i)$ deforms the $\alpha^{\rm th}$ $\Z_2$-fixed point in $T^2_{(i)}$. Terms with $\delta F_1^{(\alpha)} \delta F_2^{(\beta)} \delta F_3^{(\gamma)}$ are suppressed since $\varepsilon_{\alpha\beta\gamma}$ are not free parameters~\cite{Vafa:1994rv}. The holomorphic three-form (for $v_i \equiv 1$ ) is given by $\Omega_3 = \frac{dx_1 \wedge dx_2 \wedge dx_3}{y(x_i)}$.

For $T^6/(\Z_2 \times \Z_6^\prime)$ with discrete torsion we introduce an additional $\Z_3$ symmetry into $F_i$ and $\delta F_i^{(\alpha)}$. We then find three $\Z_3$-triplets of $\Z_2$ fixed points \textit{preserved} by $\sigma_\mathcal{R}$ with now {\it real} deformation parameters $\varepsilon^{(i)}_{1},\, \varepsilon^{(i)}_{2},\, \varepsilon^{(i)}_{3}$, and one {\it complex} triplet
 $\varepsilon^{(i)}_{4} \stackrel{ \sigma_\mathcal{R}}{=} \overline{\varepsilon}^{(i)}_{5}$ for $i \in \{1,2,3\}$. 

\section{Concrete Models on $T^6(\Z_2 \times \Z_6' \times \OR)$}

At first, we consider SO$(8)^4$ and USp$(8)^4$ models, where the D6-branes only wrap orientifold-even cycles, see~\cite{Blaszczyk:2015oia} for details on the discrete data (wrapping numbers, displacements and Wilson lines). There exists no central U(1) factor, and therefore neither D-terms in the low-energy effective action nor a potential for the deformation moduli.

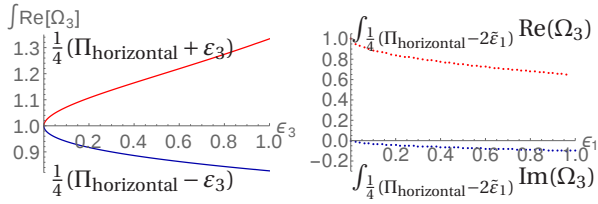
\begin{figure}
\begin{center}
\begin{tabular}{c@{\;}c}
  \includegraphics[width=.45\columnwidth]{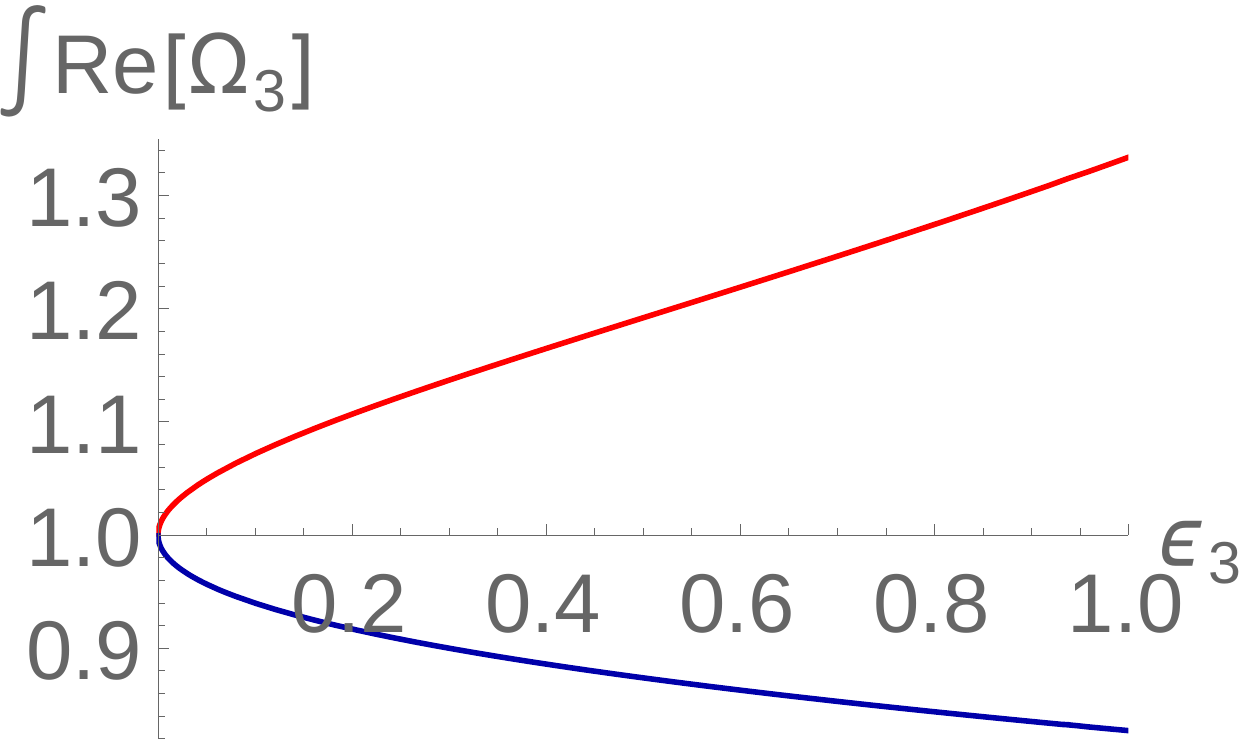}
  \begin{picture}(0,0) 
  	\put(-95,-5){$\frac{1}{4}(\Pi_\text{horizontal} - \varepsilon_3)$}
  	\put(-95,45){$\frac{1}{4}(\Pi_\text{horizontal} + \varepsilon_3)$}
  \end{picture}
  &\includegraphics[width=.45\columnwidth]{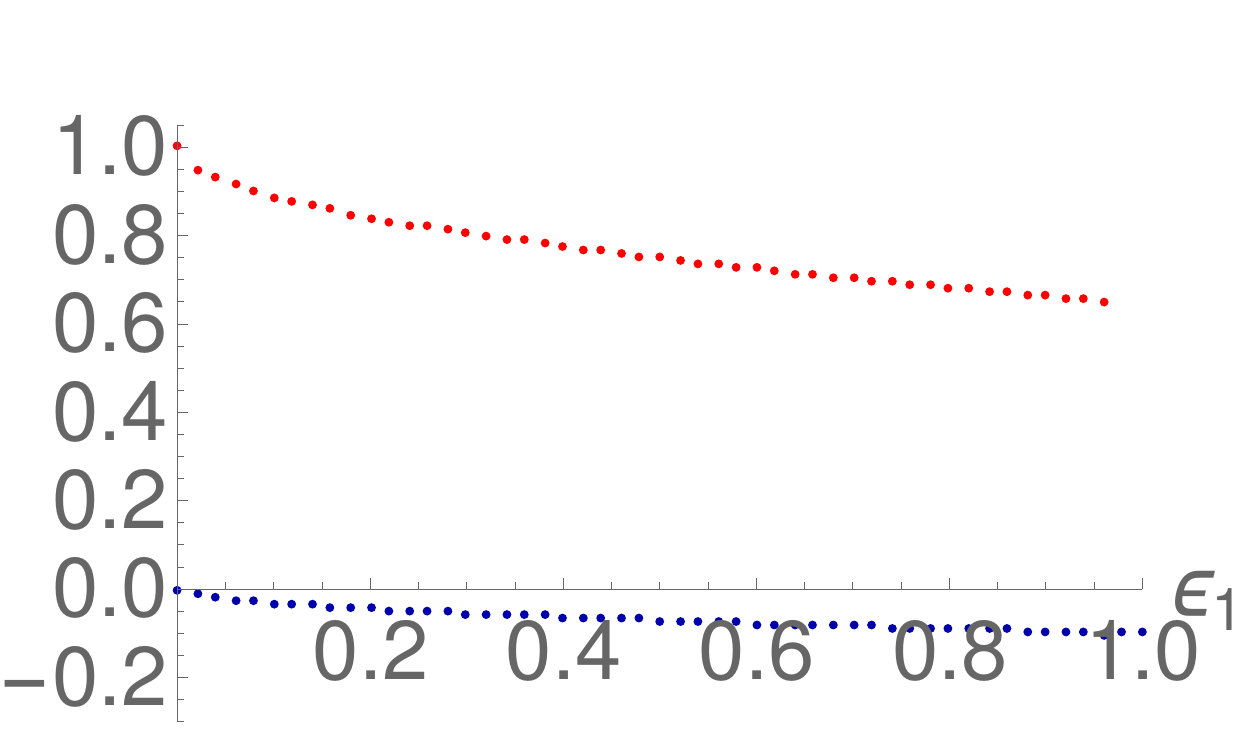}
  \begin{picture}(0,0) 
  	\put(-95,-4){$ \int_{\frac{1}{4}(\Pi_\text{horizontal} - 2 \tilde{\varepsilon}_1)} \text{Im}(\Omega_3)$}
  	\put(-95,52){$ \int_{\frac{1}{4}(\Pi_\text{horizontal} - 2 \tilde{\varepsilon}_1)} \text{Re} (\Omega_3)$}
  \end{picture} 
\end{tabular}
  \caption{\label{fig:1a} (Normalised) periods of the respective fractional {\it sLag} cycles plotted against the deformation parameter $\varepsilon_3^{(i)}$ (left) and $\varepsilon_{1}^{(i)}$ (right) for one $i \in \{1,2,3\}$.}
\end{center}
\end{figure}

The integrals over $\Re(\Omega_3)$ of a D6-brane wrapping a `horizontal' fractional cycle with exceptional contribution $\pm \boldsymbol{\varepsilon}_3^{(i)}$ are depicted in figure~\ref{fig:1a} (left), where similar graphs are obtained for the other deformation parameters $\varepsilon_1^{(i)}, \varepsilon_2^{(i)}, \varepsilon_{4+5}^{(i)}$. The sign depends on the specific D6-brane stack,
 and the square root-like dependence on $\varepsilon_3^{(i)}$ shows that the {\it sLag} property is preserved. This implies that gauge coupling constants can be varied along flat directions in the complex structure moduli space and that no stabilisation of moduli at the orbifold point occurs.

The second example is a globally defined Pati-Salam model with five stacks of D6-branes which was first presented in~\cite{Honecker:2012qr}. It has the following gauge group: \\
$\text{SU}(4)_a\times \text{SU}(2)_b\times \text{SU}(2)_c \times \text{SU}(2)_d \times \text{SU}(2)_e\times \text{U}(1)^5_{\text{massive}}$.\\
For U($N$) gauge groups, a stack of $N$ identical D-branes must wrap a cycle which is \textit{not} orientifold invariant. If a fractional D6-brane has no coupling to the orientifold-odd part of the exceptional cycle, one observes a flat direction, and the cycle stays \textit{sLag}, as in the SO$(8)^4$ and USp$(8)^4$ models. On the other hand, if there is a coupling to the orientifold-odd part, the corresponding modulus is stabilised at the supersymmetric orbifold point.
Here, the deformation breaks supersymmetry while providing a positive value for the scalar potential, and the exceptional cycle receives a non-\textit{sLag} contribution, 
i.e.\ an imaginary contribution as in figure~\ref{fig:1a} (right). Giving a \textit{vev} to the deformation modulus thus generates a Fayet-Iliopoulos term for the U(1)$\subset$U($N$) subgroup. 

The restrictions from the D6-branes limit the number of allowed deformations such that $4+4+2=10$ of the 15 deformation moduli can be stabilised at the orbifold point. In the five flat directions, the deformation $\varepsilon_3^{(3)} > 0$ can change the SU(4) coupling constant against the SU$(2)_{\rm R/L}$ couplings, while $\varepsilon_2^{(i)} \ge 0 \,, \varepsilon_1^{(3)} \ge 0 \; (i=1,2,3)$ have no influence on the low energy-effective field theory.

\section{Conclusions}

We developed an explicit description of $T^6/(\Z_2 \times \Z_{2M})$ orbifolds and their deformations in the hypersurface formalism, providing us with the technical tools to quantitatively study the effect of deformations. Therefore, we were able to directly calculate gauge coupling parameters, and we could show that these can be enhanced or diminished to adjust them in an appropriate way to fit realistic models. Furthermore, we could stabilise ten out of 15 deformation moduli in an exemplary three-generation Pati-Salam model due to the emergence of supersymmetry breaking D-terms for some U(1) factors upon moduli {\it vev}s.

\quad \\
\noindent
{\bf Acknowledgements:} 
This work is partially supported by the Cluster of Excellence PRISMA DFG no. EXC 1098, the DFG research grant HO 4166/2-1 and the GRK 1581.


\end{document}